\documentclass[pra]{revtex4}
\usepackage{amsfonts}
\usepackage{amssymb}
\usepackage{amsmath}
\usepackage{graphicx}
\usepackage[center]{subfigure}
\usepackage{color}
\usepackage{mathrsfs}

\begin{document}

\title{Two-dimensional composite solitons in Bose-Einstein condensates with
spatially confined spin-orbit coupling}
\author{Yongyao Li$^{1}$, Xiliang Zhang$^{1}$, Rongxuan Zhong$^{1}$, Zhihuan
Luo$^{2}$}
\author{Bin Liu$^{1}$, Chunqing Huang$^{1}$, Wei Pang$^{3}$}
\email{kingprotoss@gmail.com}
\author{Boris A. Malomed$^{4,1}$ }
\affiliation{$^{1}$School of Physics and Optoelectronic Engineering, Foshan University,
Foshan 528000, China \\
$^{2}$College of Electronic Engineering, South China Agricultural
University, Guangzhou 510642, China $^{3}$Department of Experiment Teaching,
Guangdong University of Technology, Guangzhou 510006, China\\
$^{4}$ Department of Physical Electronics, School of Electrical Engineering,
Faculty of Engineering, and Center for Light-Matter Interaction, Tel Aviv
University, Tel Aviv 69978, Israel}

\begin{abstract}
It was recently found that the spin-orbit (SO) coupling can help to create
stable matter-wave solitons in spinor Bose-Einstein condensates in the
two-dimensional (2D) free space. Being induced by external laser
illumination, the effective SO coupling can be applied too in a spatially
confined area. Using numerical methods and the variational approximation
(VA), we build families of 2D solitons of the semi-vortex (SV) and
mixed-mode (MM) types, and explore their stability, assuming that the
SO-coupling strength is confined in the radial direction as a Gaussian. The
most essential result is identification, by means of the VA and numerical
methods, of the minimum size of the spatial confinement for which the 2D
system maintains stable solitons of the SV and MM types.\newline
\end{abstract}

\maketitle

\section{Introduction}

Many-body self-trapping has been drawing much interest in studies of atomic
Bose-Einstein condensates (BECs). In particular, creation of stable two- and
three-dimensional (2D and 3D) solitons is a challenging issue, as the usual
cubic self-attraction destabilizes all formally available multidimensional
solitons due to the possibility of the collapse \cite{Berge1998,Sulem1999}.
Two schemes were theoretically elaborated to solve the stability problem for
matter-wave solitons in the 2D and 3D free space. One is the use of nonlocal
nonlinearity, which may be induced by the Van der Waals interactions between
Rydberg atoms \cite{Pohl2011}, dipole-dipole interactions between atoms or
molecules carrying magnetic or electric dipolar moments \cite%
{Santos2006,Tikhonenkov2008,Raykuang2017}, or the microwave-mediated local
field effect in spinor BECs \cite{Jqin2016,accelerating}. The second scheme
relies upon the use of beyond-mean-field corrections, induced by quantum
fluctuations, which are represented by the Lee-Huang-Yang (LHY) terms added
to the underlying Gross-Pitaevskii equations (GPEs). The latter approach has
made it possible to theoretically predict \cite{Petrov2015}-\cite{QDSOC2018}
and experimentally create self-trapped ``quantum droplets", in dipolar \cite%
{Schmitt2016,Barbut2016,Chomaz2016} and binary BECs \cite%
{Cabrera2018,Cabrera2,Inguscio,2Dvortex,CuiX2018}.

Recently, an unexpected result was reported, predicting a possibility to
create absolutely stable (ground-state) and metastable matter-wave solitons
in the 2D \cite{HS2014} and 3D \cite{Zhang2015} free space, respectively,
with the help of the spin-orbit (SO) coupling, which can be induced in
binary (pseudo-spinor) BEC by means of appropriate laser fields, see
original works \cite{Nature}-\cite{Anderson} and reviews \cite{NatureRev}-%
\cite{Zhai}. While a majority of experimental works aimed to create the SO
coupling in effectively 1D settings, an experimental realization of an
effectively 2D SO coupling was reported too \cite%
{2D-experiment,2D-experiment2}. In the setting considered in Ref. \cite%
{HS2014}, the SO coupling can protect 2D solitons against collapsing,
creating a ground state \cite{Sherman}, which is otherwise missing in 2D
GPEs with the cubic self-attraction \cite{Dias2015}-\cite{Guihua}. The
collapse remains possible in the presence of SO coupling, starting with the
norm of the input which exceeds the threshold value for the onset of the 2D
collapse.\ Similar settings can be implemented in optics, predicting the
creation of spatiotemporal solitons (``light bullets") in planar dual-core
waveguides and twisted cylinder waveguide with the self-focusing Kerr
nonlinearity, respectively \cite{YVK2015,HS2016,Haohuang}. Further, the
interplay between the SO coupling and anisotropic dipole-dipole interactions
in 2D free space can create stripe solitons \cite{Xuyong2015}, solitary
vortices \cite{Xunda2016,Bingjin1,Shimei,Bingjin2}, and gap solitons \cite%
{gapsoliton2017} (2D free-space gap solitons can also be created in
SO-coupled BECs with contact interactions, at appropriate values of
parameters \cite{HS2018}). Recently, it was also found that the combination
of LHY and SO-coupling terms in 2D creates anisotropic ``quantum droplets"
in spinor BECs \cite{QDSOC2017}.

Previous works on 2D and 3D solitons in SO-coupled BECs tacitly assumed that
the SO-couplings was applied homogeneously in the entire space. Because this
effect is engineered by applied laser fields, it can be applied in a
spatially confined area. This possibility was analyzed, in the framework of
the 1D SO-coupling model, in Ref. \cite{confined-1D}. While stable 1D
matter-wave solitons can be created without the use of the SO coupling \cite%
{Randy}-\cite{Weiman}, \cite{Luca}, the analysis reported in Ref. \cite%
{confined-1D} has revealed new possibilities, such as the creation of stable
two-soliton bound states. The purpose of the present work is to construct 2D
solitons supported by spatially confined SO coupling, which is a challenging
issue, as 2D solitons are unstable without the SO coupling. Thus, in
particular, a relevant problem is to identify the minimum area carrying the
SO coupling which is necessary to maintain the solitons' stability. We
address this problem, assuming an isotropic shape of the spatial modulation
of the local SO strength, with a Gaussian dependence on the radial
coordinate. The results are obtained by means of an analytical variational
approximation (VA)\ and systematic numerical calculations. The rest of the
paper is structured as follows: the model and VA are introduced in Sections
II and III, respectively, and numerical results, including their comparison
with predictions of the VA are summarized in Section IV. The paper is
concluded by Sec. V.

\section{The model}

As said above, we consider the binary BECs, with a pseudo-spinor wave
function $(\phi _{+},\phi _{-})$, whose components are SO-coupled in a
finite 2D area. The mean-field model of this system is based on the
Lagrangian,
\begin{equation}
\mathbb{L}=\int \int \mathscr{L}dxdy,  \label{L}
\end{equation}%
\begin{gather}
\mathscr{L}=-\frac{i}{2}\left( \phi _{+}^{\ast }\frac{\partial \phi _{+}}{%
\partial t}+\phi _{-}^{\ast }\frac{\partial \phi _{-}}{\partial t}+\mathrm{%
c.c.}\right)  \notag \\
+\frac{1}{2}\left( \left\vert \nabla \phi _{+}\right\vert ^{2}+\left\vert
\nabla \phi _{-}\right\vert ^{2}\right) -\frac{1}{2}\left( |\phi
_{+}|^{4}+|\phi _{-}|^{4}\right) -\gamma |\phi _{+}|^{2}|\phi _{-}|^{2}
\notag \\
+\frac{\lambda (r)}{2}\left\{ \left[ \phi _{+}^{\ast }\frac{\partial \phi
_{-}}{\partial x}-\phi _{-}^{\ast }\frac{\partial \phi _{+}}{\partial x}%
-i\left( \phi _{+}^{\ast }\frac{\partial \phi _{-}}{\partial y}+\phi
_{-}^{\ast }\frac{\partial \phi _{+}}{\partial y}\right) \right] +\mathrm{%
c.c.}\right\} .  \label{density}
\end{gather}%
where $\mathrm{c.c.}$ stands for the complex conjugate expression. The SO
coupling of the Rashba type is accepted here, with a strength confined to
values of the radial coordinate $r\lesssim L$:%
\begin{equation}
\lambda (r)=\lambda _{0}\exp \left( -r^{2}/L^{2}\right) ,  \label{lambda}
\end{equation}%
where and $\lambda _{0}\equiv 1$ may be fixed by means of rescaling.
Further, $\gamma $ is the relative strength of the cross attraction, while
the strength of the self-attraction is normalized to be $1$. The Hamiltonian
corresponding to Lagrangian (\ref{L}) is%
\begin{gather}
E=\int \int \left( \mathcal{E}_{\mathrm{K}}+\mathcal{E}_{\mathrm{N}}+%
\mathcal{E}_{\mathrm{SOC}}\right) dxdy,  \notag \\
\mathcal{E}_{\mathrm{K}}=\frac{1}{2}\left( \left\vert \nabla
u_{+}\right\vert ^{2}+\left\vert \nabla u_{-}\right\vert ^{2}\right) ,~%
\mathcal{E}_{\mathrm{N}}=-\frac{1}{2}\left[ \left(
|u_{+}|^{4}+|u_{-}|^{4}\right) -2\gamma |u_{+}|^{2}|u_{-}|^{2}\right] ,
\notag \\
\mathcal{E}_{\mathrm{SOC}}=\frac{\lambda (r)}{2}\left\{ \left[ u_{+}^{\ast
}\left( \frac{\partial u_{-}}{\partial x}-i\frac{\partial u_{-}}{\partial y}%
\right) -u_{-}^{\ast }\left( \frac{\partial u_{+}}{\partial x}+i\frac{%
\partial u_{+}}{\partial y}\right) \right] +\mathrm{c.c.}\right\} ,
\label{E}
\end{gather}
where $\mathcal{E}_{\mathrm{K,N,SOC}}$ are densities of kinetic,
interaction, and SO-coupling energies, respectively.

The GPE system is derived from Lagrangian (\ref{L}) as the Euler-Lagrange
equations, written here in polar coordinates $\left( r,\theta \right) $, as
suggested by the fact that $\lambda $ is defined as a function of $r$ in Eq.
(\ref{lambda}) [the following relations are useful is this context: $%
\partial_{x}-i\partial_{y}=e^{-i\theta}(\partial_{r}-ir^{-1}\partial_{%
\theta}),
\partial_{x}+i\partial_{y}=e^{i\theta}(\partial_{r}+ir^{-1}\partial_{\theta})
$]

\begin{eqnarray}
i\frac{\partial \phi _{+}}{\partial t} &=&-\frac{1}{2}\nabla ^{2}\phi
_{+}-(|\phi _{+}|^{2}+\gamma |\phi _{-}|^{2})\phi _{+}+\lambda
(r)e^{-i\theta }\left( \frac{\partial \phi _{-}}{\partial r}-\frac{i}{r}%
\frac{\partial \phi _{-}}{\partial \theta }\right) +\frac{1}{2}e^{-i\theta }%
\frac{d\lambda }{dr}\phi _{-},  \notag \\
i\frac{\partial \phi _{-}}{\partial t} &=&-\frac{1}{2}\nabla ^{2}\phi
_{-}-(|\phi _{-}|^{2}+\gamma |\phi _{+}|^{2})\phi _{-}-\lambda (r)e^{i\theta
}\left( \frac{\partial \phi _{+}}{\partial r}+\frac{i}{r}\frac{\partial \phi
_{+}}{\partial \theta }\right) -\frac{1}{2}e^{i\theta }\frac{d\lambda }{dr}%
\phi _{+}.  \label{GPE2}
\end{eqnarray}%
%
%
%
%
%
%
%
%
%
%
Note that the last terms in Eq. (\ref{GPE2}), produced by the $r$-dependence
of $\lambda $, may be considered as a specific form of the Rabi coupling.

Stationary solutions to Eq. (\ref{GPE2}) with chemical potential $\mu $ are
looked for as
\begin{equation}
\left\{ \phi _{\pm }\left( x,y,t\right) \right\} =e^{-i\mu t}u_{\pm }\left(
x,y\right) ,  \label{u}
\end{equation}%
where functions $u_{\pm }$ satisfy equations
\begin{eqnarray}
\mu u_{+} &=&-\frac{1}{2}\nabla ^{2}u_{+}-(|u_{+}|^{2}+\gamma
|u_{-}|^{2})u_{+}+\lambda (r)e^{-i\theta }\left( \frac{\partial u_{-}}{%
\partial r}-\frac{i}{r}\frac{\partial u_{-}}{\partial \theta }\right) +\frac{%
1}{2}e^{-i\theta }\frac{d\lambda }{dr}u_{-},  \notag \\
\mu u_{-} &=&-\frac{1}{2}\nabla ^{2}u_{-}-(|u_{-}|^{2}+\gamma
|u_{+}|^{2})u_{-}-\lambda (r)e^{i\theta }\left( \frac{\partial u_{+}}{%
\partial r}+\frac{i}{r}\frac{\partial u_{+}}{\partial \theta }\right) -\frac{%
1}{2}e^{i\theta }\frac{d\lambda }{dr}u_{+},  \label{uu}
\end{eqnarray}%
which can be derived from their own Lagrangian density:%
\begin{gather}
\mathscr{L}_{\mathrm{stat}}=-\mu \left( \left\vert u_{+}\right\vert
^{2}+|u_{-}|^{2}\right) +\frac{1}{2}\left( \left\vert \nabla
u_{+}\right\vert ^{2}+\left\vert \nabla u_{-}\right\vert ^{2}\right) -\frac{1%
}{2}\left( |u_{+}|^{4}+|u_{-}|^{4}\right) -\gamma |u_{+}|^{2}|u_{-}|^{2}
\notag \\
+\frac{\lambda (r)}{2}\left\{ \left[ e^{-i\theta }u_{+}^{\ast }\left( \frac{%
\partial u_{-}}{\partial r}-\frac{i}{r}\frac{\partial u_{-}}{\partial \theta
}\right) -e^{i\theta }u_{-}^{\ast }\left( \frac{\partial u_{+}}{\partial r}+%
\frac{i}{r}\frac{\partial u_{+}}{\partial \theta }\right) \right] +\mathrm{%
c.c.}\right\} .  \label{u density}
\end{gather}

\section{Semi-vortices (SVs) and the variational approximation (VA) for them}

Equations (\ref{uu}) admit solutions in the form of a \textit{semi-vortex}
(SV):%
\begin{equation}
u_{+}=f(r),~u_{-}=\exp \left( i\theta \right) rg(r),  \label{SV}
\end{equation}%
with $\mu <0$. This ansatz is exactly compatible with Eq. (\ref{uu}), but
real functions $f(r)$ and $g(r)$ must be found numerically. They
exponentially decay $\sim \exp \left( -\sqrt{-2\mu }r\right) $ at $%
r\rightarrow \infty $ [note that $\lambda (r)$ vanishes at $r\rightarrow
\infty $, hence the SO-coupling does not affect the asymptotic form at $%
r\rightarrow \infty $], and take finite values, $f(r=0)\neq 0$ and $%
g(r=0)\neq 0$, at $r=0$, with zero values of the derivatives: $f^{\prime
}(r=0)=g^{\prime }(r=0)=0$.

The SV may be approximated by the Gaussian variational ansatz, with
different amplitudes, $A$ and $B$, and common width $W$, cf. Ref. \cite%
{HS2014}:%
\begin{equation}
u_{+}(r)=A\exp \left( -\frac{r^{2}}{2W^{2}}\right) ,~u_{-}(r)=Br\exp \left(
i\theta -\frac{r^{2}}{2W^{2}}\right) .  \label{ansatz}
\end{equation}%
The substitution of this ansatz in Lagrangian density (\ref{u density}) and
spatial integration yields the effective Lagrangian corresponding to the
ansatz:%
\begin{equation}
\frac{\mathbb{L}}{\pi }=-\mu \left( A^{2}W^{2}+B^{2}W^{4}\right) +\frac{A^{2}%
}{2}+B^{2}W^{2}-\frac{A^{4}W^{2}}{4}-\frac{B^{4}W^{6}}{8}-\frac{\gamma
A^{2}B^{2}W^{4}}{4}+\frac{2ABL^{2}W^{2}}{L^{2}+W^{2}},  \label{Leff}
\end{equation}%
which gives rise to the variational equations, $\partial \mathbb{L}/\partial
A=\partial \mathbb{L}/\partial B=\partial \mathbb{L}/\partial \left(
W^{2}\right) =0$, i.e.,%
\begin{eqnarray}
&&\frac{2L^{2}BW^{2}}{L^{2}+W^{2}}=2\mu AW^{2}-A+A^{3}W^{2}+\frac{\gamma }{2}%
AB^{2}W^{4},  \notag \\
&&\frac{2L^{2}A}{L^{2}+W^{2}}=2\mu BW^{2}-2B+\frac{1}{2}B^{3}W^{4}+\frac{%
\gamma }{2}A^{2}BW^{2},  \notag \\
&&\frac{2L^{4}AB}{\left( L^{2}+W^{2}\right) ^{2}}=\mu A^{2}+2\mu
B^{2}W^{2}-B^{2}+\frac{1}{4}A^{4}+\frac{3}{8}B^{4}W^{4}+\frac{\gamma }{2}%
A^{2}B^{2}W^{2}.  \label{eqns}
\end{eqnarray}%
The total norm of ansatz (\ref{ansatz}) is%
\begin{equation*}
N=\int \left[ |\phi _{+}(\mathbf{r})|^{2}+|\phi _{-}(\mathbf{r})|^{2}\right]
d\mathbf{r\equiv }N_{+}+N_{-}
\end{equation*}%
\begin{equation}
=\pi \left( A^{2}W^{2}+B^{2}W^{4}\right) .  \label{N}
\end{equation}%
In particular, analysis of Eqs. (\ref{eqns}) and (\ref{N}) reproduces the
known fact \cite{HS2014} that, in the uniform space ($L=\infty $), SVs exist
with norms falling below a limit value, $N<N_{T}$, where $N_{T}$ is the norm
of the \textit{Townes' soliton} \cite{Townes,Berge1998,Sulem1999} produced
by the single GPE in the 2D setting. The present version of the VA predicts
the known approximate value, $N_{T}^{(\mathrm{VA})}=2\pi $ \cite{Lisak}, a
numerically exact one being $N_{T}\approx 5.85$ \ref{LcrSV}. Results
produced by the VA for VSs are compared to numerical findings in the next
section. In particular, the VA predicts the minimum size of the SO-coupling
area, $L_{\mathrm{cr}}$, necessary for supporting 2D solitons.

\section{Numerical results}

\subsection{Stationary semi-vortices (SVs) and mixed modes (MMs)}

\begin{figure}[t]
\subfigure[]{\includegraphics[width=0.3\columnwidth]{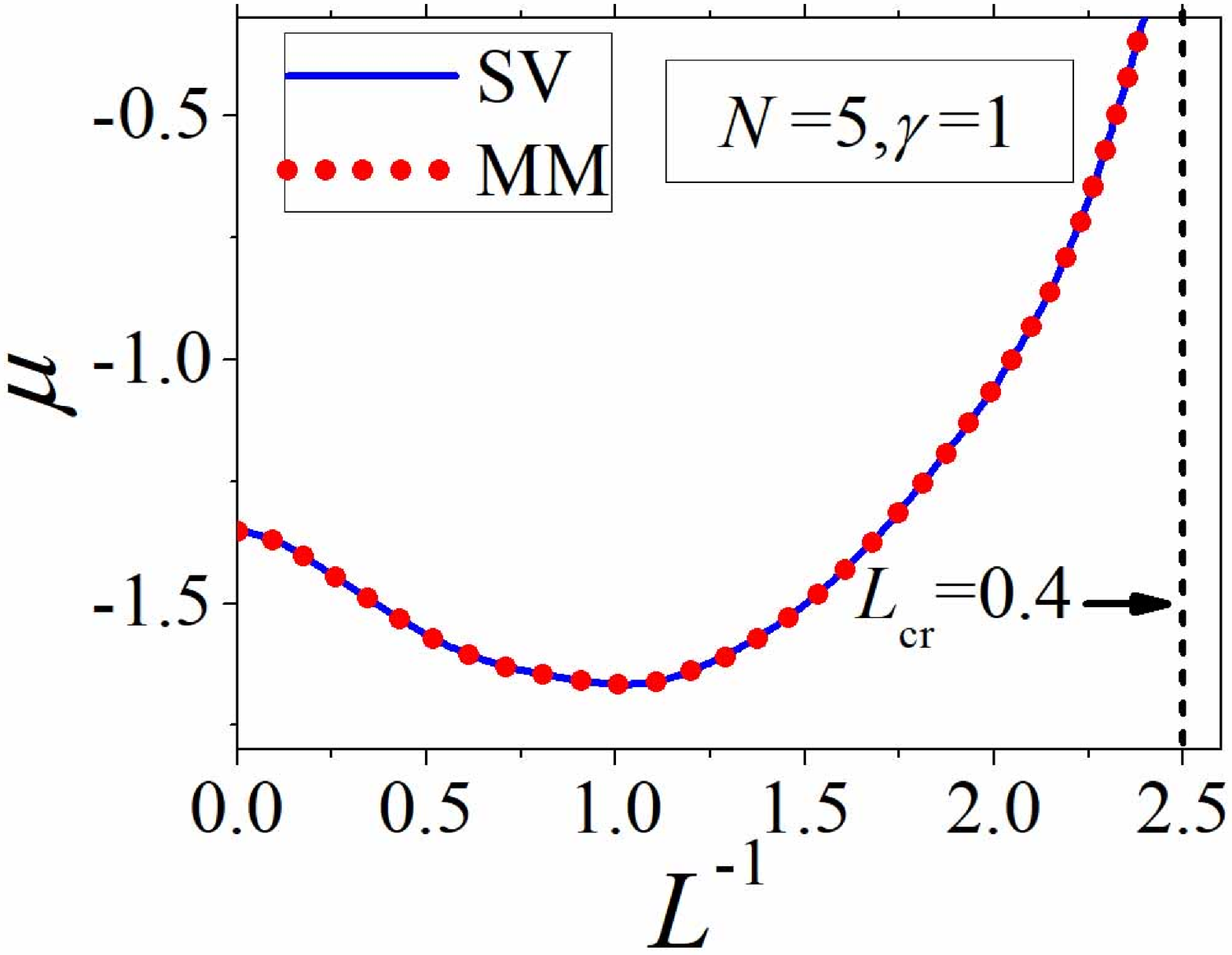}} %
\subfigure[]{\includegraphics[width=0.3\columnwidth]{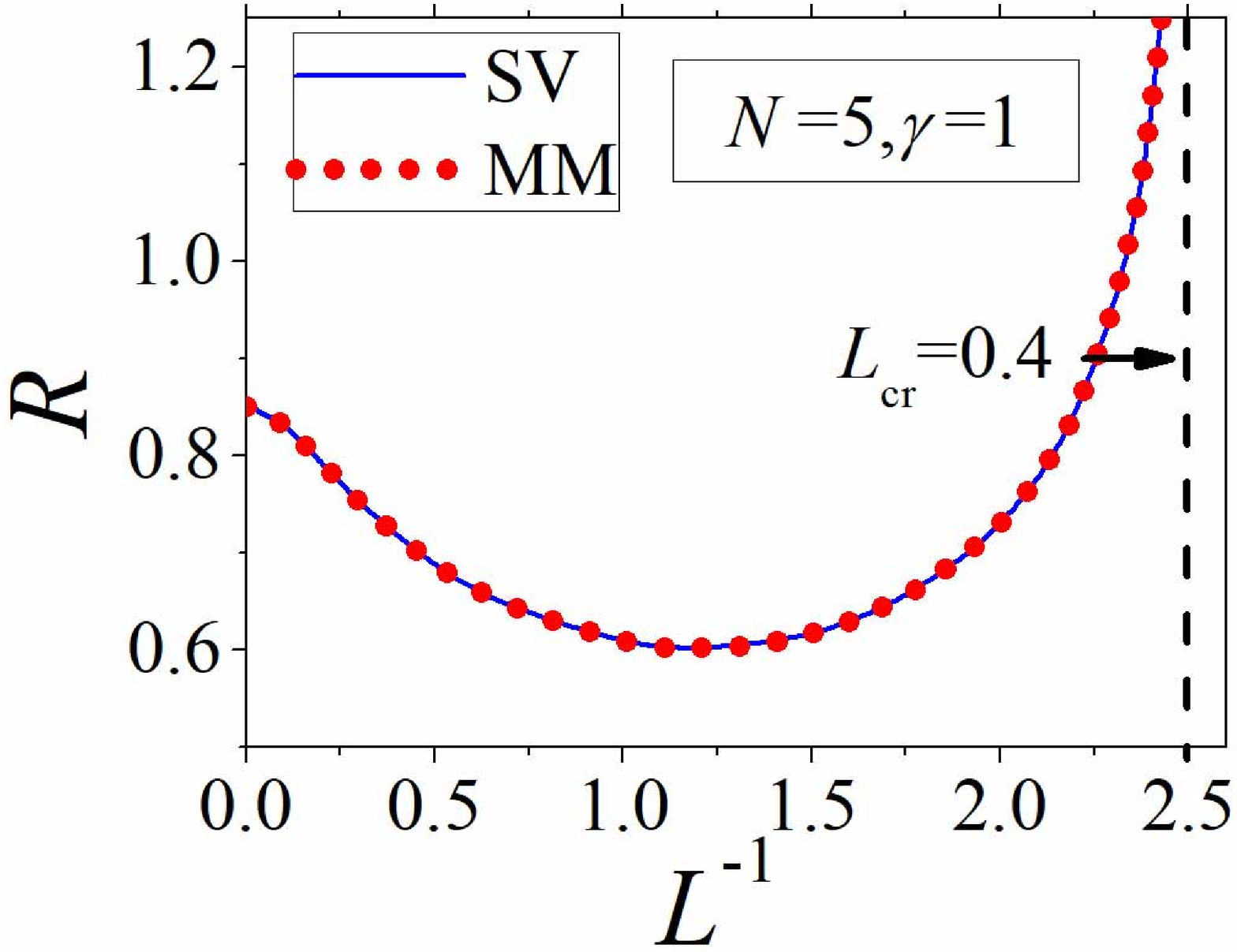}} %
\subfigure[]{\includegraphics[width=0.3\columnwidth]{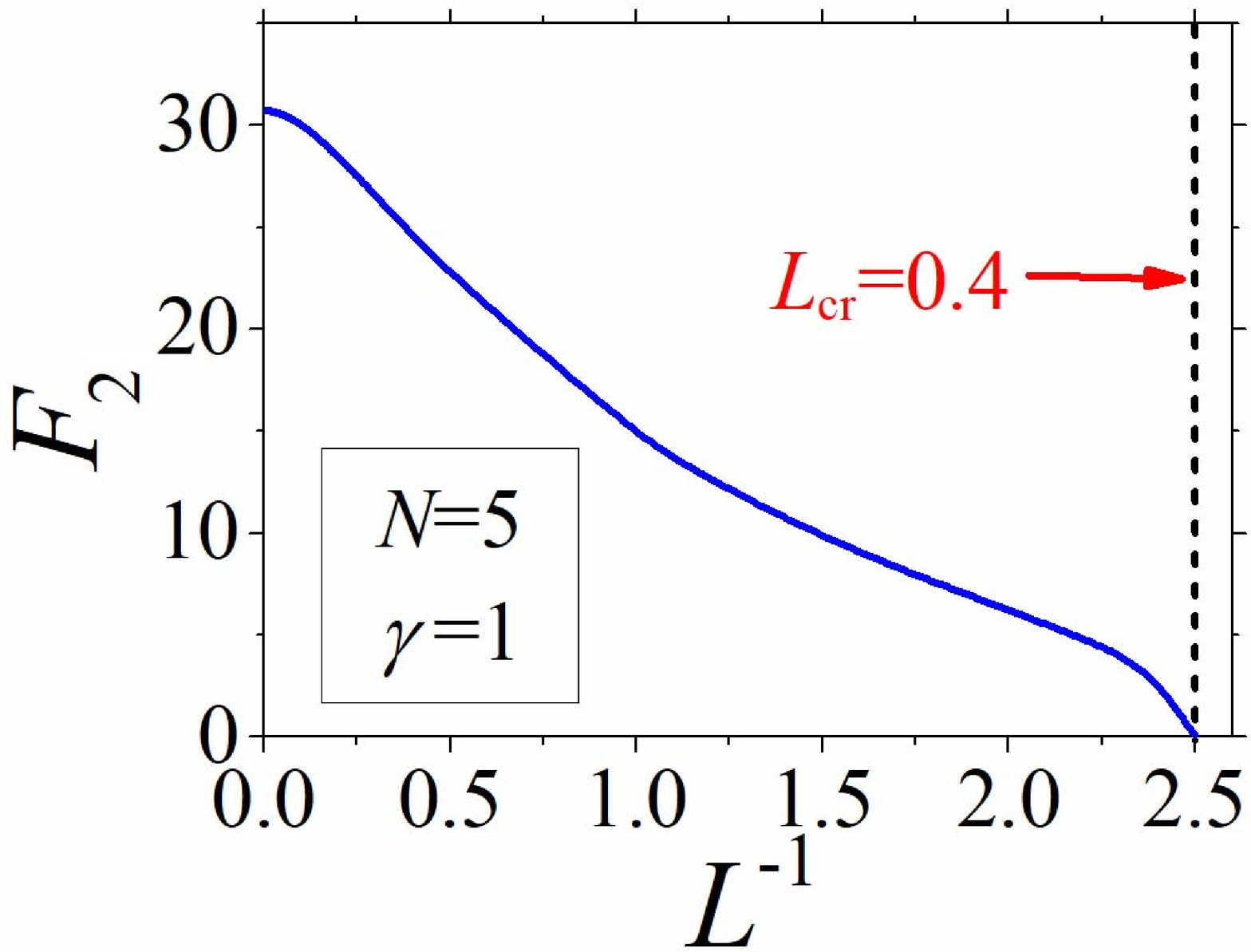}}
\caption{(Color online) (a,b) Effective radius (\protect\ref{RR}) and the
chemical potential of the SVs (blue solid curves) MMs (red dot curves)
versus $L$. Here we fix $(N,\protect\gamma )=(5,1)$. (c)The
vorticity-carrying norm share $F_{2}$ of SVs [see Eq. (\protect\ref{F2})]
vs. $L$. }
\label{Char}
\end{figure}

\begin{figure}[t]
{\includegraphics[width=0.7\columnwidth]{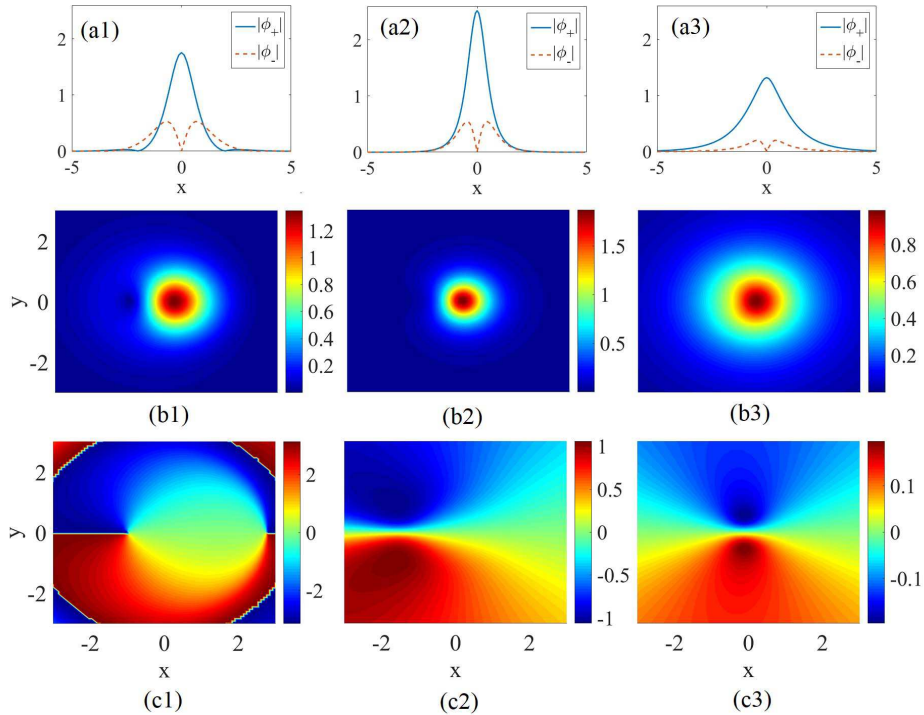}}
\caption{(Color online) (a1-a3) Amplitude profiles in cross-sections of the
fundamental and vortex components (blue solid and red dashed curves,
respectively) of stable SVs. (b1-b3) 2D amplitude pattern of component $%
\protect\phi_{+}$ of stable MMs. (c1-c3) The phase patterns of $\protect\phi%
_{+}$ corresponding to panels (b1-b3), respectively. The size of the
SO-coupling confinement, $L$, from left to right columns is $L=\infty $, $1$%
, $0.41$, respectively. Other parameters are $\protect\gamma = 1$ and $N=5$.}
\label{MM}
\end{figure}

\begin{figure}[t]
\subfigure[]{\includegraphics[width=0.4\columnwidth]{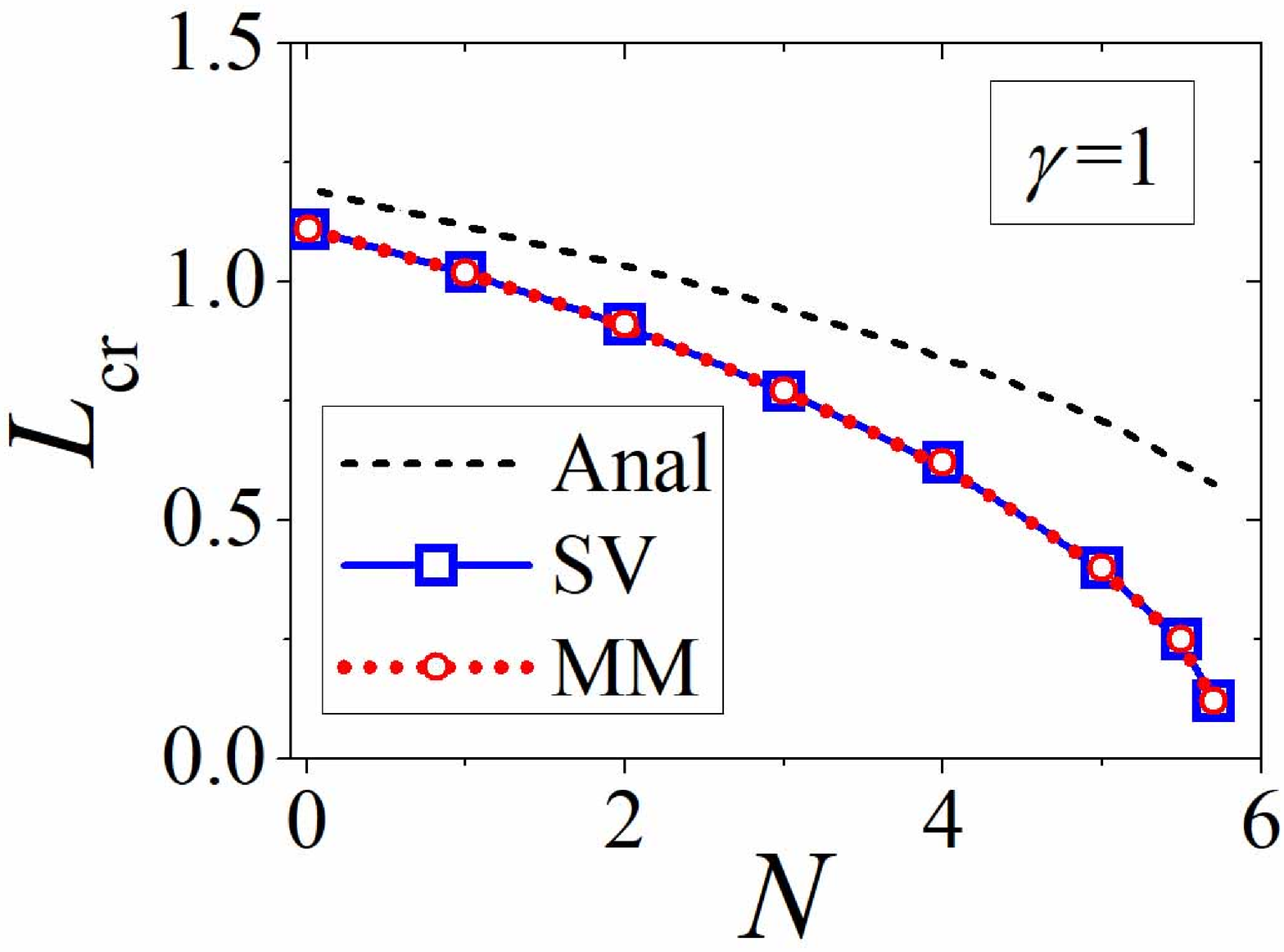}} %
\subfigure[]{\includegraphics[width=0.4\columnwidth]{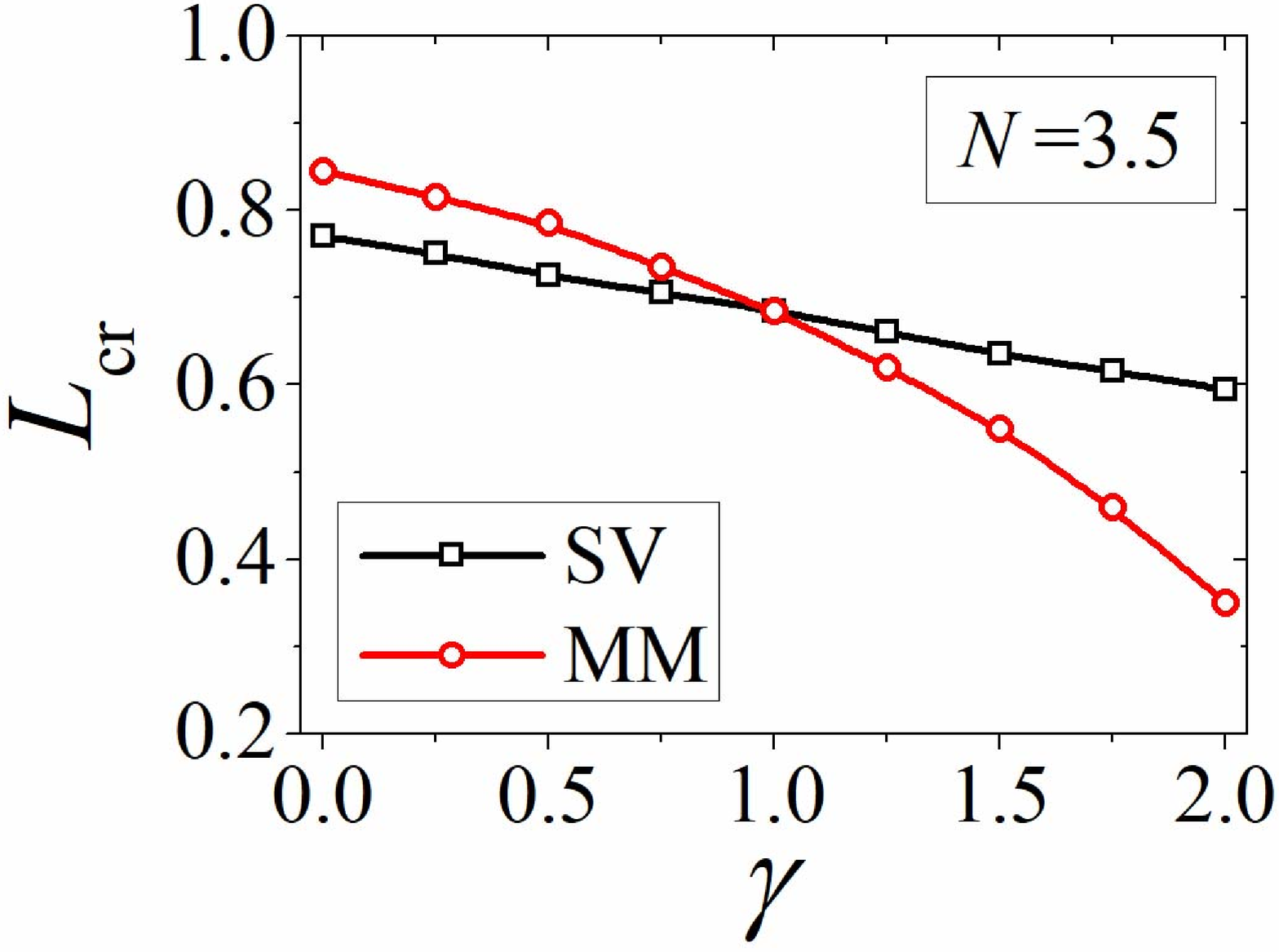}} %
\subfigure[]{\includegraphics[width=0.32\columnwidth]{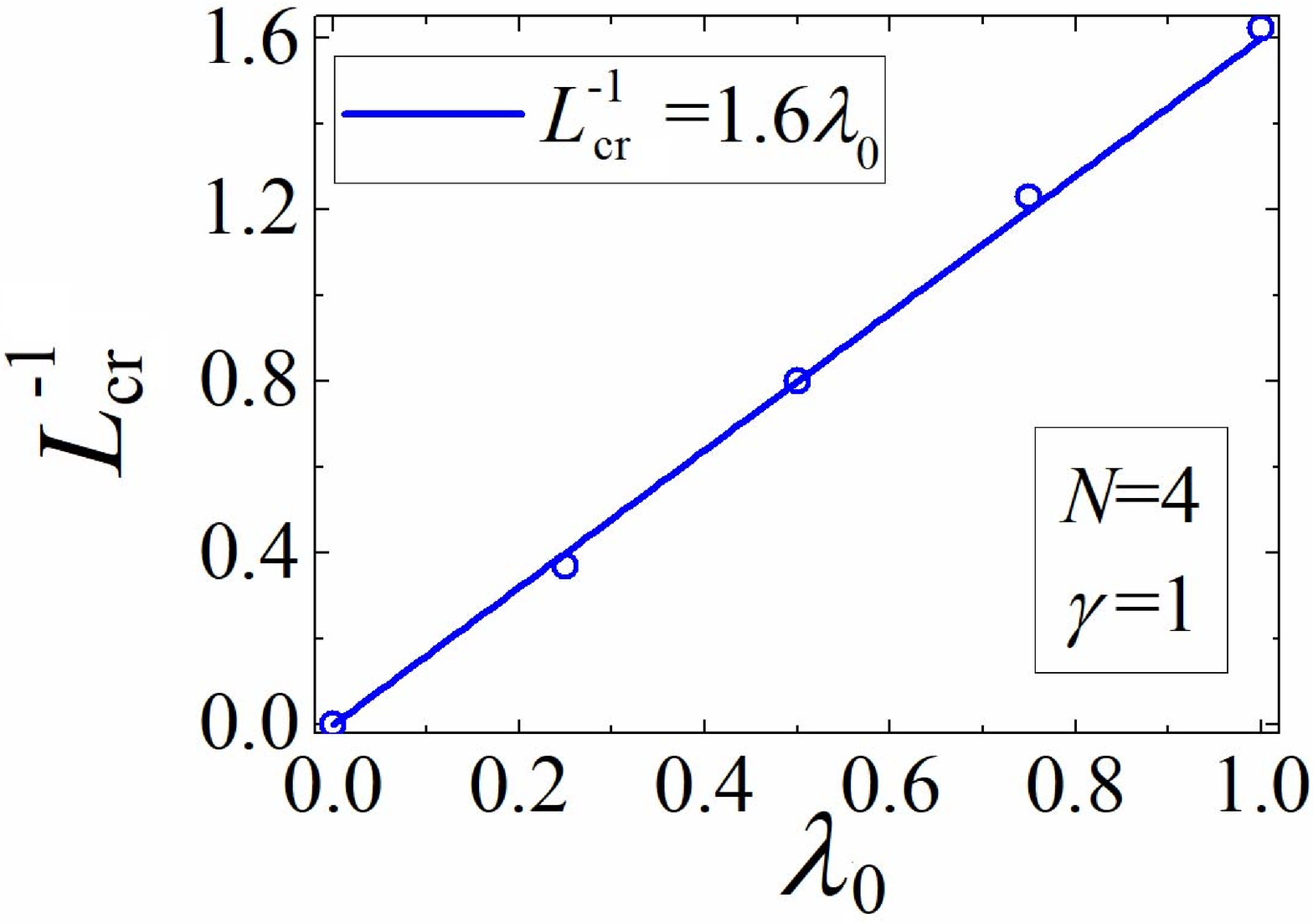}}
\caption{(Color online) (a) $L_{\mathrm{cr}}$ for SVs (blue solid with
squares) and MMs (red dots with circles) vs. $N$ at $\protect\gamma =1$. The
dashed curve is dependence $L_{\mathrm{cr}}(N)$ predicted by the VA, see the
text. (b) $L_{\mathrm{cr}}$ of SVs (black squares) and MMs (red circles) vs.
$\protect\gamma $ at $N=3.5$. (c) $L_{\mathrm{cr}}$ for SVs vs. $\protect%
\lambda_{0}$, at fixed $\protect\gamma=1$ and $N=4$. }
\label{LcrSV}
\end{figure}

\begin{figure}[t]
\subfigure[]{\includegraphics[width=0.4\columnwidth]{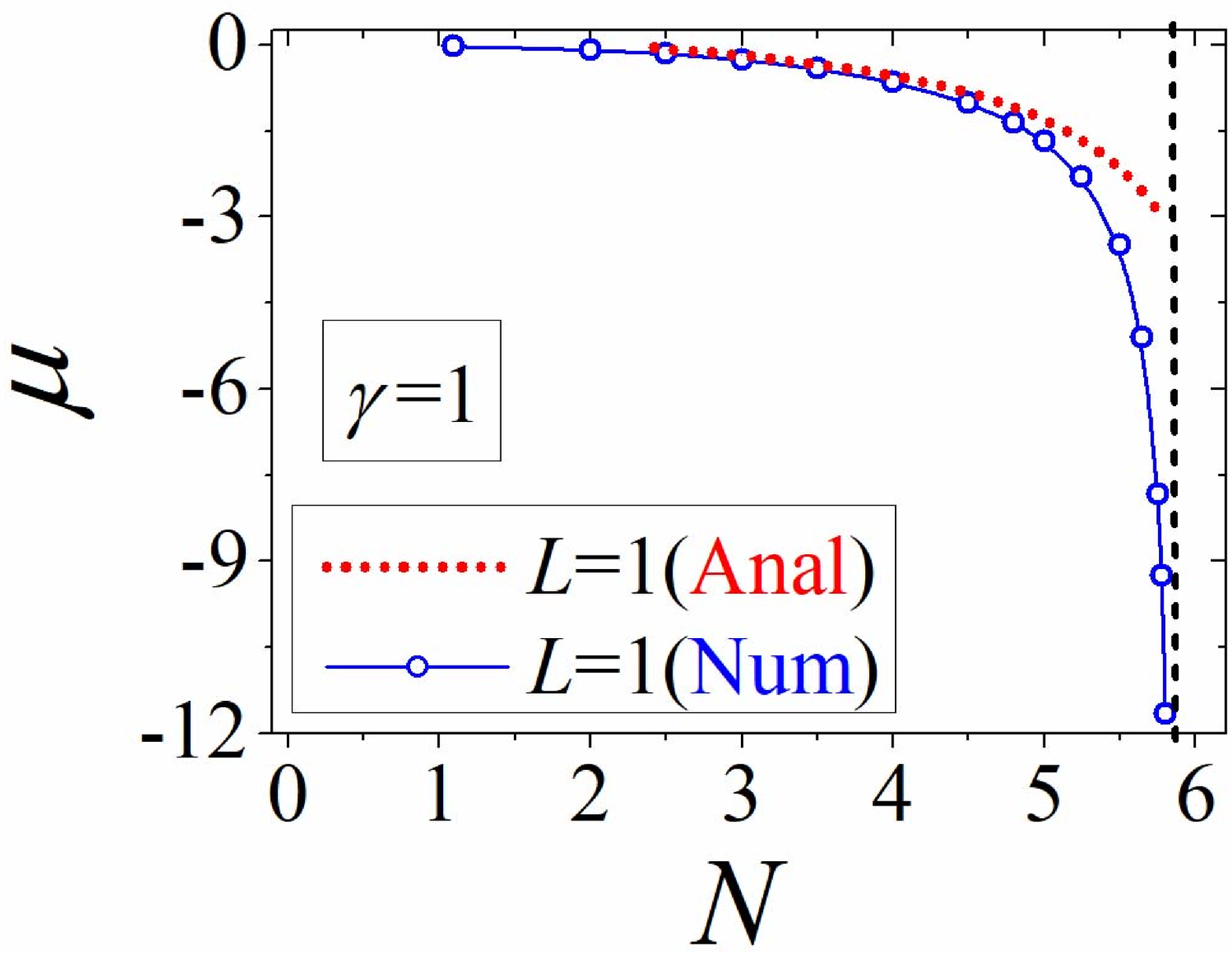}} %
\subfigure[]{\includegraphics[width=0.4\columnwidth]{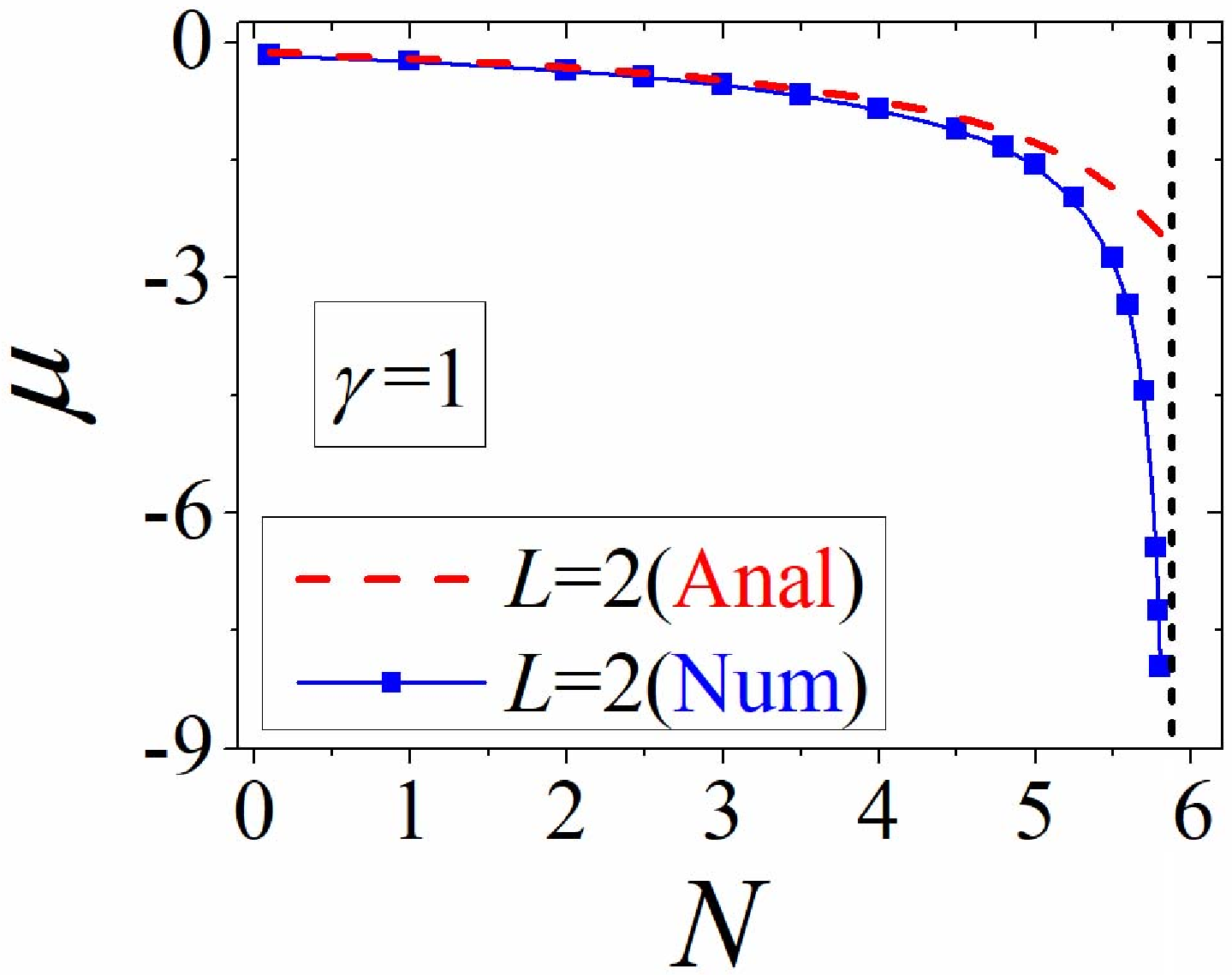}} %
\subfigure[]{\includegraphics[width=0.4\columnwidth]{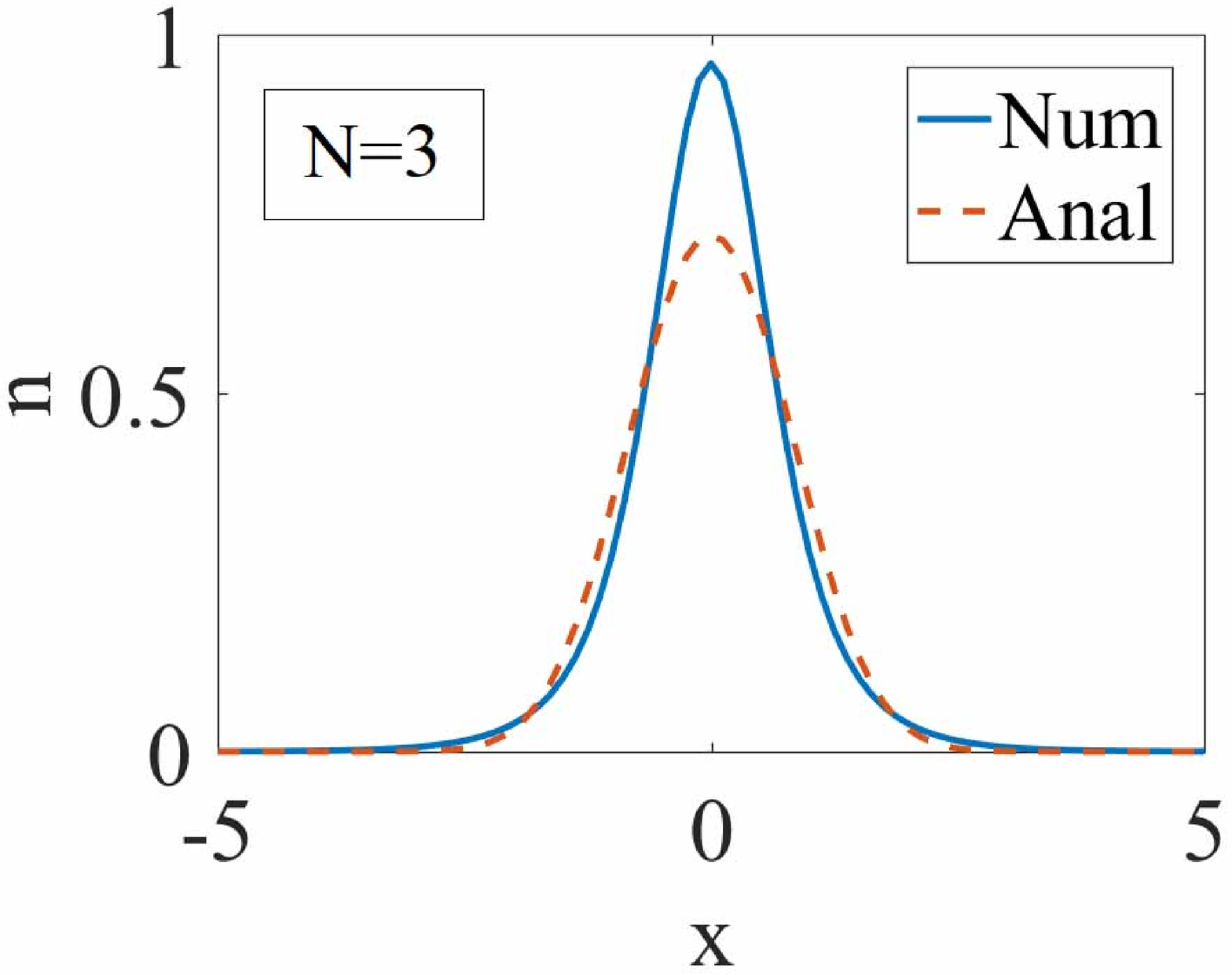}} %
\subfigure[]{\includegraphics[width=0.4\columnwidth]{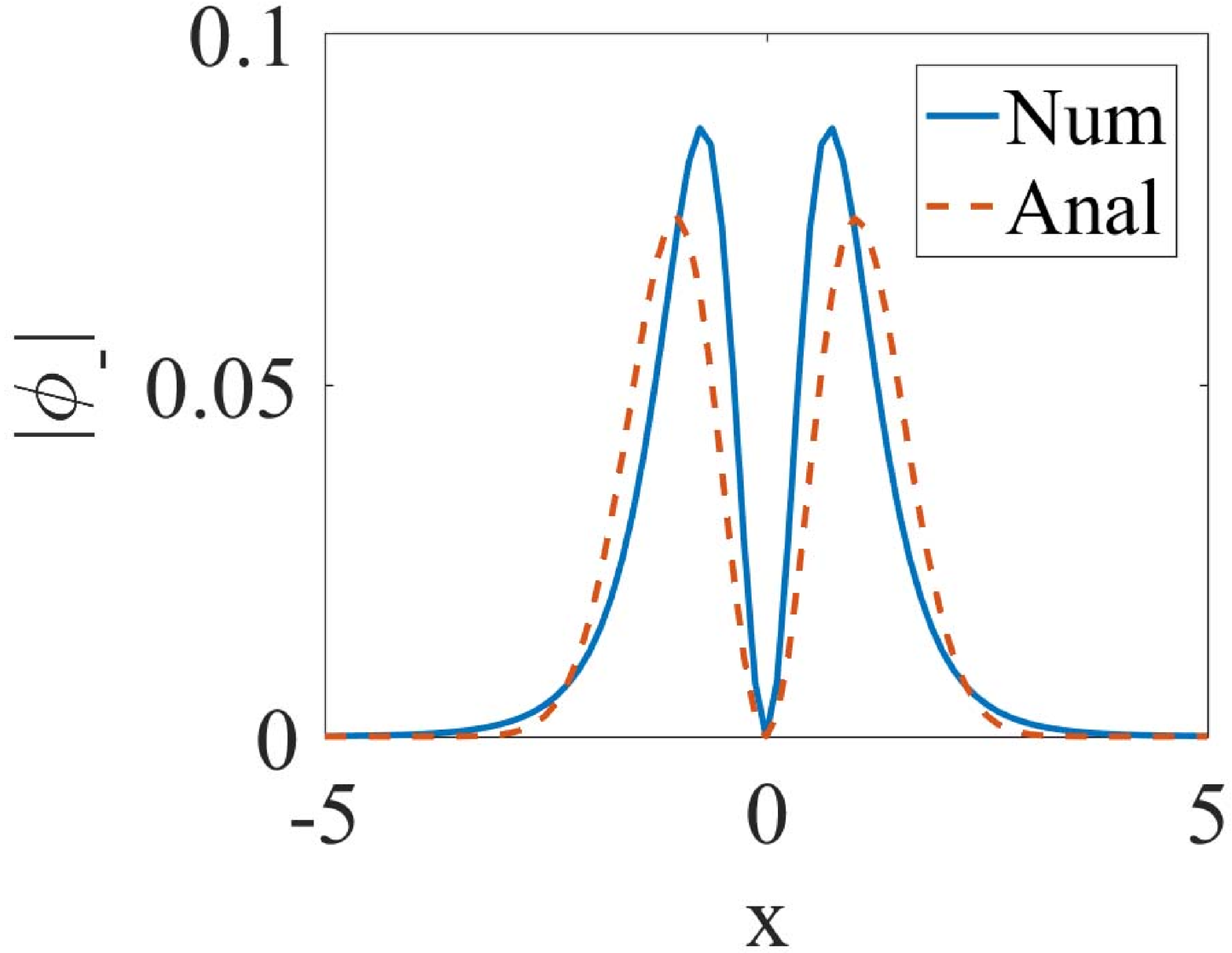}}
\caption{(Color online) (a,b) Chemical potential $\protect\mu $ of the
numerically generated (``Num") SVs and MMs, as a function of $N$, for $L=1$
(a) and $2$ (b), respectively, at $\protect\gamma =1$ (the values of $%
\protect\mu $ fully coincide for the SVs and MMs). The red dot and dashed
curves display the quasi-analytical (``Anal") predictions of the VA for the
same values, $L=1$ and $2$, in panels (a) and (b), respectively, which are
generated by Eqs. (\protect\ref{eqns}) and (\protect\ref{N}). (c,d) The
comparison between the numerical result (the blue solid line) and the its VA
counterpart (the red dashed line) for the cross section of the total density
profile, $n(\mathbf{r})$, and the absolute value of the vortex component, $|%
\protect\phi _{-}(\mathbf{r})|$, for the SV soliton with $(N,\protect\gamma %
,L)=(3,1,1)$.}
\label{muP}
\end{figure}

According to Ref. \cite{HS2014}, two types of 2D solitons, the
above-mentioned SVs and mixed modes (MMs), can be produced by the SO-coupled
GPEs. It is relevant to mention that, in the uniform space ($L=\infty $),
the MMs exist with the norm falling below the limit value,%
\begin{equation}
N<N_{\lim }=2\left( 1+\gamma \right) ^{-1}N_{T},  \label{lim}
\end{equation}%
where $N_{T}$ is the above-mentioned norm of the Townes' soliton, which sets
the limit for the SV's norm.

Stationary SVs can be numerically obtained, solving Eq. (\ref{GPE2}) by
means of the imaginary-time method \cite{Chiofalo2000,JKyang2008}, starting from
input
\begin{equation}
\phi _{+}^{(0)}=A_{+}\exp (-\alpha _{+}r^{2}),\quad \phi
_{-}^{(0)}=A_{-}r\exp (i\theta -\alpha _{-}r^{2}),  \label{VAguess}
\end{equation}%
with real constants $A_{\pm }$ and $\alpha _{\pm }>0$. Note that this input
is similar to, but different from variational ansatz (\ref{ansatz}).
Similarly, MMs are produced by the imaginary-time integration initiated by
input
\begin{equation}
\phi _{\pm }^{(0)}=A_{1}\exp (-\alpha _{1}r^{2})\mp A_{2}r\exp (-\alpha
_{2}r^{2}\mp i\theta ),  \label{MMguess}
\end{equation}%
with $\alpha _{1,2}>0$. The imaginary-time integration method, initiated by
these two inputs, converges, respectively, to soliton solutions of the SV
and MM types. Unlike Eq. (\ref{SV}), an ansatz built in the form of Eq. (\ref%
{MMguess}) is not compatible with Eq. (\ref{uu}). Nevertheless, the general
structure represented by the ansatz, i.e., a superposition of vorticities $%
(0,+1)$ and $(0,-1)$ in the two components. is also featured by numerical
solutions for the MM.

To address effects of confinement size $L$ of the SO coupling, which is
defined in Eq. (\ref{lambda}), we define an effective radius of the soliton,
as
\begin{equation}
R=\left( {\frac{\int r^{2}n(\mathbf{r})d\mathbf{r}}{\int n(\mathbf{r})d%
\mathbf{r}}}\right) ^{1/2},  \label{RR}
\end{equation}%
where $n(\mathbf{r})=|\phi _{+}(\mathbf{r})|^{2}+|\phi _{-}(\mathbf{r})|^{2}$
is the total density of the solution. For the SVs, it also relevant to
define the relative share of the total number of atoms which are kept in the
vortex component:
\begin{equation}
F_{2}={\frac{N_{-}}{N}}\times 100\%,  \label{F2}
\end{equation}
as per definition of $N_{-}$ given by Eq. (\ref{N}). For MMs solutions,
norms of their components are always equal.
Dependences of these characteristics on $L$, obtained from numerical
solutions, are produced below, along with results verifying stability of the
solitons.

Figures \ref{Char}(a,b) display the chemical potentials and radii of the SVs
and MMs, defined by Eq. (\ref{RR}), for characteristic values of other
parameters, $(N,\gamma )=(5,1)$, as functions of the SO-coupling confinement
size, $L$. Note that the values of $\mu $ and $R$ for SVs and MMs are
identical for $\gamma =1$, which is a manifestation of the specific
degeneracy of the soliton families in this case (in the uniform space, with $%
L=\infty $, the SVs and MMs are limit cases of a broader soliton family with
an additional intrinsic parameter; the same may be true in the case of
finite $L$, which should be a subject for additional analysis). Values of $%
\mu (L)$ and $R(L)$ decrease with $L$ varying from infinity to $L\approx
0.83 $, and then increase with the subsequent decrease of $L$. This behavior
implies that, initially, the solitons undergo self-compression with the
reduction of the size of the SO-coupling area, which is changed by
expansion. As $L$ approaches the critical value, $L_{\mathrm{cr}}\approx 0.4$%
, at which the solitons suffer delocalization, $R(L)$ asymptotically
diverges, while $\mu (L)$ vanishes in the same limit. Solitons do not exist
at $L<L_{\mathrm{cr}}$. Further, \ref{Char}(c) shows that the share of the
total norm in the vortex component of the SV monotonously decay with the
decrease of $L$, vanishing in the limit of $L=L_{\mathrm{cr}}$. A similar
trend occurs for the MMs, in both components of which the vortex terms are
vanishing at $L\rightarrow L_{\mathrm{cr}}$.

Figure \ref{MM} shows typical examples of stable SVs, as well as the
amplitude and phase patterns of stable MMs, at different values of $L$. It
is observed that the decrease of $L $ makes the MM's shape more circular,
which is a natural consequence of squeezing the mode by the spatial
confinement. As concerns SVs, due to their axial symmetry they are displayed
by means of the radial cross sections.

The critical size $L_{\mathrm{cr}}$ being the most essential characteristic
of the present setting, we display its dependence on $N$ and $\gamma $ in
Fig. \ref{LcrSV}. In particular, Fig. \ref{LcrSV}(a) shows comparison of the
VA-predicted and numerically found curves $L_{\mathrm{cr}}(N)$ for SVs. The
VA predicts $L_{\mathrm{cr}}(N)$ as the smallest value of $L$ for which,
with given $N$, numerical solution of variational equations (\ref{eqns}) and

(\ref{N}) generates a meaningful solution for parameters $A$, $B$, and $W$.
It is seen that the agreement is reasonable, the numerically generated SVs
being somewhat more robust, as $L_{\mathrm{cr}}$ is slightly smaller for
them.

Further, the identical equality of the values of $L_{\mathrm{cr}}$ for SVs
and MMs at $\gamma =1$, observed in Fig. \ref{LcrSV}(a), is a
straightforward corollary of Eq. (\ref{lim}): in the limit of $L\rightarrow
L_{\mathrm{cr}}$, the vortex terms in the MM vanish, and this soliton
degenerates into a bound states of two Townes' solitons, which gives rise to
the expression for its norm obtained by means of rescaling (\ref{lim}) from $%
N_{T}$. Then, $M_{\lim }$ is identical to $N_{T}$ in the case of $\gamma =1$%
. Furthermore, the same argument suggests that, for equal values of $L_{%
\mathrm{cr}}$ and given $\gamma $, the respective limit values of $N$, at
which $L=L_{\mathrm{cr}}$ is attained by the SVs and MMs are related
similarly to Eq. (\ref{lim}):%
\begin{equation}
N_{\lim }^{(\mathrm{MM})}(L_{\mathrm{cr}})=2\left( 1+\gamma \right)
^{-1}N_{\lim }^{(\mathrm{SV})}(L_{\mathrm{cr}}),  \label{limlim}
\end{equation}%
which is corroborated by numerical data. It is worthy to note that,
according to Eq. (\ref{limlim}) $L_{\mathrm{cr}}$ for SVs and MMs with equal
norms are different at $\gamma \neq 1$. In particular, in Fig. \ref{LcrSV}%
(b) we display the $L_{\mathrm{cr}}(\gamma )$ dependences for the two
soliton species, which agree with the prediction of Eq. (\ref{limlim}). This
panel also shows that $L_{\mathrm{cr}}$ of both species decrease with the
increase of $\gamma $.

The decrease of $L_{\mathrm{cr}}$ with the increase of $N$ and $\gamma $,
clearly seen in Fig. \ref{LcrSV}, is a natural trend, as the stronger
nonlinearity, corresponding to larger $N$ and/or $\gamma $, leads to
self-compression of the solitons, making them less sensitive to the the
spatial confinement of the SO coupling. Inverting dependence $L_{\mathrm{cr}%
}(N)$, displayed in Fig. \ref{LcrSV}(a), i.e., considering it as $N(L)$, one
can interpret it in an alternative way: for given $L$, the SVs and MMs
exist, severally, in regions%
\begin{equation}
N_{\mathrm{\lim }}^{(\mathrm{SV})}(L)<N<N_{T},~N_{\mathrm{\lim }}^{(\mathrm{%
MM})}(L)<N<2\left( 1+\gamma \right) ^{-1}N_{T},  \label{NN}
\end{equation}%
while in the case of $L=\infty $ there is no lower norm threshold necessary
for the existence of stable SVs and MMs \cite{HS2014}.

The fact that $L_{\mathrm{cr}}$, i.e., the localization size of the wave
functions, remains finite at $N\rightarrow 0$ in Fig. \ref{LcrSV}(a)
demonstrates that the spatially localized SO coupling plays the role of an
effective trapping potential in the linear system. A similar effect was
mentioned in Ref. \cite{confined-1D}, where a 1D localized potential was
induced by a finite area of SO-coupling.

A dependence between $L_{\mathrm{cr}}$ and $\lambda _{0}$ (the strength of
SO coupling) was addressed too. Figure \ref{LcrSV}(c) shows $L_{\mathrm{cr}%
}^{-1}$ as a function of $\lambda _{0}$, for SV states. The figure shows
that $L_{\mathrm{cr}}^{-1}$ vanishes almost linearly at $\lambda
_{0}\rightarrow 0$, in the interval of $\lambda _{0}\in \lbrack 0,1]$. This
linear dependence can be qualitatively explained by noting that, if the SV
state with norm $N$ fills a 2D area of size $L$, the respective squared
amplitude can be estimated as $A^{2}\sim N/L^{2}$. The SOC terms may balance
the self-attractive nonlinearity as long as the corresponding relation
holds, $\lambda _{0}/L\sim A^{2}\sim N/L^{2}$, An obvious corrollary of the
latter estimate is $L^{-1}\sim \lambda _{0}$, in agreement with Fig. \ref%
{LcrSV}(c).

For fixed values of $L$ and $\gamma $, soliton families are naturally
characterized by dependences $\mu (N)$, which are displayed for $L=1$ and $2$
with $\gamma =1$ in Fig. \ref{muP}(a,b), respectively. An essential fact is
that $\mu (N)$ curves satisfy the Vakhitov-Kolokolov criterion, $d\mu /dN<0$%
, which is a well-known necessary stability condition for the solitons \cite%
{VK,Berge1998,Sulem1999}. Moreover, the comparison between the numerical
results and the VA-predicted dependence $\mu (N)$ for SVs, see Eq. (\ref%
{eqns}), shows that they coincide very well for small values of $N$,
deviating at larger $N$, the reason being that the simple ansatz (\ref%
{ansatz}) is not accurate enough for large norms. In addition, the
comparison between typical numerically found shapes of the SV and the
respective VA prediction is shown in Fig. \ref{muP}(b,c), showing
qualitative agreement.

\subsection{Stability of the 2D solitons}

\begin{figure}[t]
\subfigure[]{\includegraphics[width=0.4\columnwidth]{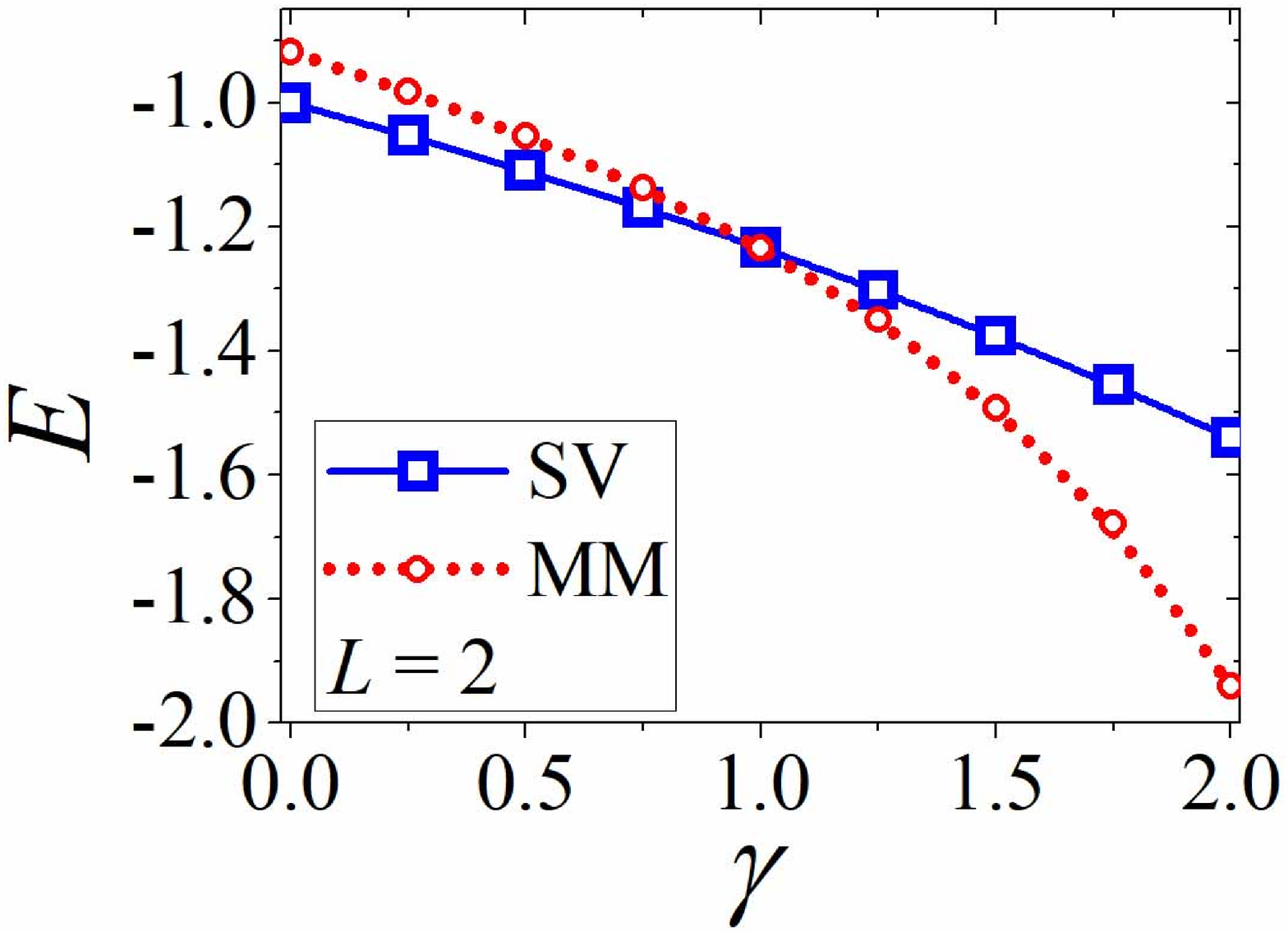}}%
\subfigure[]{\includegraphics[width=0.4\columnwidth]{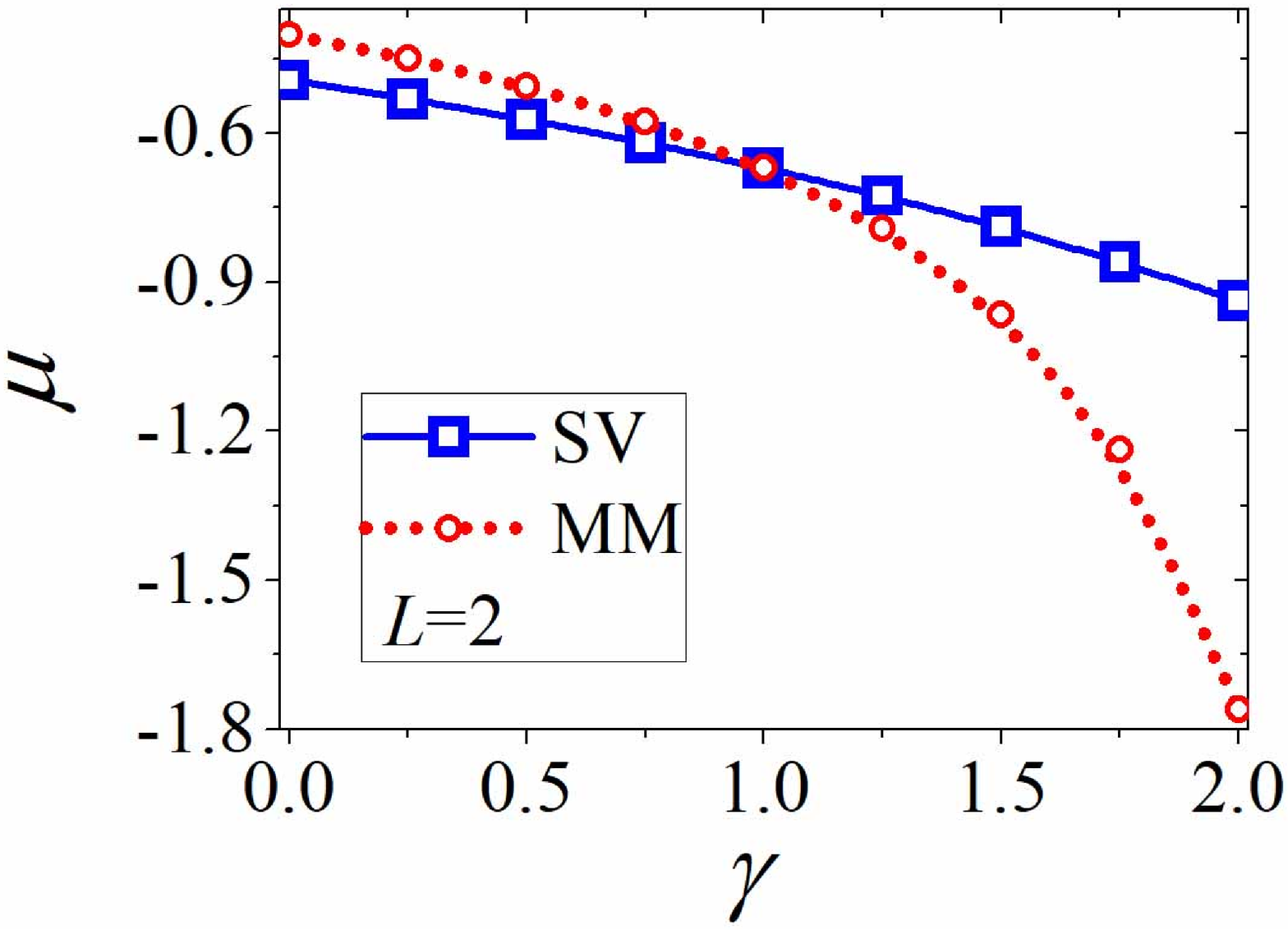}}
\caption{(Color online) (a,b) The energy and chemical potential of SVs (blue
solid with squares) and MMs (red dots with circles) vs. $\protect\gamma $
with $(N.L)=(3.5,2)$. These two panels indicate that the SV and MM
degenerate at $\protect\gamma =1$, the condition of a Manakov\textexclamdown
\={}%
s type. . }
\label{energy}
\end{figure}
\begin{figure}[t]
\subfigure[]{\includegraphics[width=0.9\columnwidth]{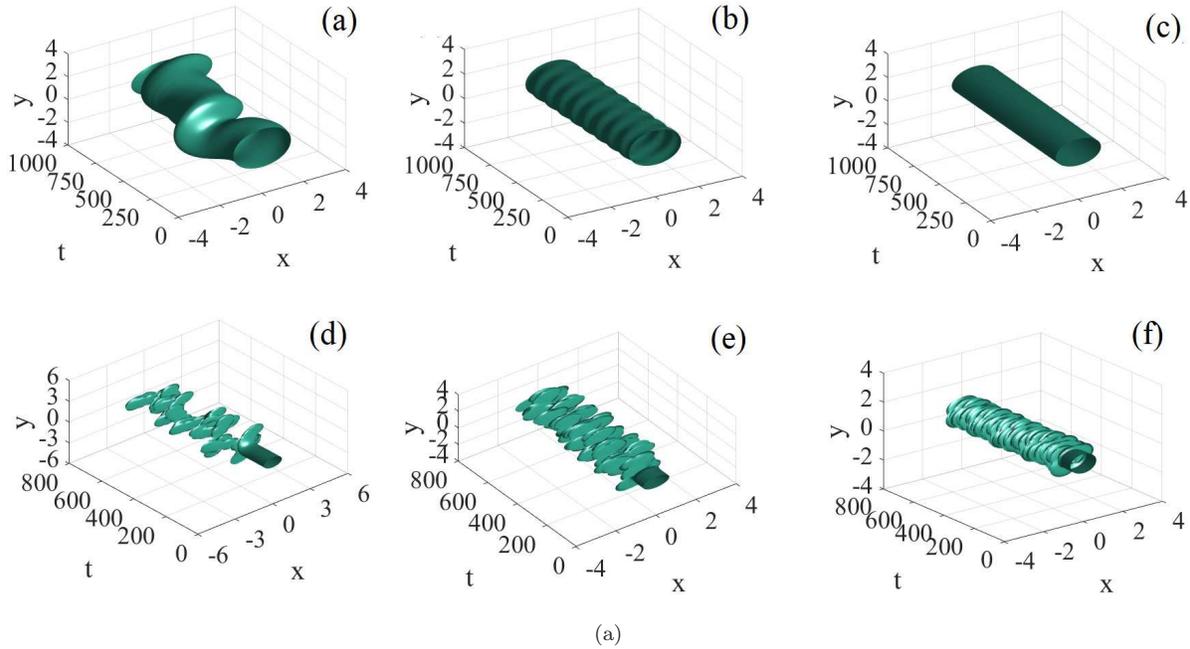}} %
\caption{(Color online) (a,b,c) Simulations of the perturbed evolution of
unstable MMs (in the case when they do not represent the energy minimum),
shown by means of the density profile, $n(\mathbf{r},t)$, for $(N,\protect%
\gamma ,L)=(3.5,0,20)$ (a), $(3.5,0,10)$ (b), and $(3.5,0,2)$ (c). (d,e,f)
The same for unstable SVs, for $(N,\protect\gamma ,L)=(3.5,2,10)$ (d), $%
(3.5,2,5)$ (e) and $(3.5,2,2)$ (f). }
\label{Unstablepro}
\end{figure}

In Ref. \cite{HS2014} it was found that, in the uniform space ($L=\infty $),
the SVs and MMs are stable, respectively, at $\gamma \geq 1$ and $\gamma
\geq 1$, where they realize the ground state of the system, i.e., the energy
minimum for given $N$. At $\gamma >1$, the SVs, whose energy exceeds that of
the MMs, are subject to weak instability, which sets them in spontaneous
motion. Similarly, the MMs are unstable at $\gamma <1$, where they tend to
spontaneously rearrange into SVs, with lower energy.

In the present system, with $L<\infty $, the ground-state switch between SVs
and MMs also happens. Fig. \ref{energy}(a) shows the energies of the SVs and
MMs with $(N,L)=(3.5,2)$ as a function of $\gamma$. It is seen that the SV
and MM realize the energy minimum, which are always stable, at $\gamma >1$
and $\gamma <1$, respectively. Their energies are equal to each other at $%
\gamma=1$, which is the system of the Manakov's type \cite{Manakov1993}. For
the comparison's sake, $\mu (\gamma )$ curves for the same parameters are
displayed in Fig. \ref{energy}(b), showing that $\mu _{\mathrm{SV}}=\mu _{%
\mathrm{MM}}$ point is also at $\gamma =1$, in accordance with the
above-mentioned degeneracy of the soliton families in this case, cf. Fig. %
\ref{muP}(a). Note that both the ground states and ones different from them
are produced here by the imaginary-time-integration method. In this
connection, it is relevant to mention that non-ground states in SO-coupled
systems were previously produced by means of the imaginary-time integration,
provided that the input and integration procedure are subject to specific
constraints, and the numerical algorithm is precise enough, to prevent a
spontaneous transition to the ground states, see Refs. \cite{QDSOC2017},
\cite{HS2014}, \cite{HS2016}, and \cite{Chunqing2018,Rongxuan2018}.

Similar to the situation for $L=\infty $, explored in Ref. \cite{HS2014},
the solitons which do not correspond to the energy minimum tend to become
unstable. However, in the system with finite $L$ the instability, which
includes spontaneous drift of the solitons, may be partly suppressed by the
confinement. To illustrate the results, Fig. \ref{Unstablepro} displays
simulated evolution of the density profiles for unstable MMs at $\gamma =0$,
and unstable SVs at $\gamma =2$, at different values of $L$. It is seen that
their drift is indeed confined by the finite values of $L$. Actually, the
confinement may effectively suppress the MM's instability, as seen in Fig. %
\ref{Unstablepro}(c), or transform the MM into a robust breather, see Fig. %
\ref{Unstablepro}(b). For the SVs which do not correspond to the energy
minimum, the instability remains conspicuous even in the presence of the
relatively tight spatial confinement.

Lastly, in addition to the fundamental 2D solitons considered above, the
SO-coupled system can also produce excited states \cite%
{HS2014,Chunqing2018,Rongxuan2018}, which are produced by adding the same
vorticity, $S\geq 1$, to both components of the 2D soliton. In particular,
excited states of SVs can be generated by input $\phi _{+}^{\mathrm{(SV)}%
}=A_{+}r^{S}\exp (-\alpha _{+}r^{2}+iS\theta )$,$~\phi _{-}^{\mathrm{(SV)}%
}=A_{-}r^{S}\exp (-\alpha _{-}r^{2}+i\left( S+1\right) \theta )$, where $%
A_{\pm }$ and $\alpha _{\pm }>0$ are real constants. Numerical simulations
demonstrate that all the excited states are unstable in the present model
too.

\section{Conclusion}

The objective of this work is to study the shapes and stability of 2D
solitons of the SV (semi-vortex) and MM (mixed-mode) in the self-attractive
pseudo-spinor BEC,\ with SO coupling applied in a confined area, following
the analysis of effects of the spatial confinement in the 1D system \cite%
{confined-1D}. Using numerical methods and the variational approximation, we
have found that, with the decrease of the confinement radius, $L$, profiles
shrink at first, and then expand to infinity (with the amplitude decaying to
zero) when $L$ approaches the critical value, $L_{\mathrm{cr}}$, below which
2D solitons do not exist. The dependences of $L_{\mathrm{cr}}$ on the
solitons' norm, $N$, and the relative strength of the cross-attraction, $%
\gamma $, are produced, on the basis of numerical results, $L_{\mathrm{cr}}$
being smaller for stronger nonlinearity, i.e., larger $N$ and $\gamma $. In
addition to the stability of the solitons which play the role of the ground
state, i.e., SV at $\gamma <1$ and MM at $\gamma >1$, unstable MMs (which do
not represent the ground state) may be partly stabilized by the spatial
confinement of the SO coupling.

As an extension of the present work, a challenging possibility is to address
3D solitons in the binary BEC\ with a spatially confined strength of the SO
coupling, following the analysis for the 3D uniform space developed in Ref.
\cite{Zhang2015}.

\begin{acknowledgments}
We appreciate valuable discussions with G. J\={u}zeliunas, Y. V. Kartashov,
and V. V. Konotop, and assistance in numerical calculations provided by Hao
Huang. This work was supported, in part, by NNSFC (China) through grants No.
11874112,11575063, by the joint program in physics between NSF and
Binational (US-Israel) Science Foundation through project No. 2015616, and
by the Natural Science Foundation of Guangdong Province, through grant No.
2015A030313639. B.A.M. appreciates a foreign-expert grant from the Guangdong
province (China).
\end{acknowledgments}

\end{document}